\documentclass[noshowpacs,amsmath,
twocolumn,
superscriptaddress,
8pt,
aps,prb]{revtex4-2}
\usepackage{caption}
\usepackage{subcaption}
\usepackage{lineno}

\usepackage{setspace}
\usepackage{amsmath}
\usepackage{multirow}
\usepackage{graphicx}

\usepackage{verbatim}
\usepackage{amsfonts}
\usepackage{amssymb}
\usepackage{epstopdf}
\usepackage{dcolumn}
\usepackage{xcolor}
\usepackage{verbatim}
\usepackage[colorlinks=true, frenchlinks=false, urlcolor=blue, linkcolor=blue, citecolor=blue]{hyperref}
\usepackage{siunitx}
\usepackage[version=4]{mhchem}
\usepackage{textcomp}
\setcitestyle{super} 
\DeclareGraphicsExtensions{.pdf,.eps,.png,.jpg,.mps}

\begin{document}

\title{Overcoming the noise-tracking-bandwidth limits in Free-running Dual-Comb Interferometry}

\author{Wei Long$^{1,2,*}$, Yujia Ji$^{1,2,3,4}$, Xiangze Ma$^{1,2,3,4}$, Dijun Chen$^{1,2,3,4,*}$\\
\medskip
$^1${Wangzhijiang Innovation Center for Laser, Aerospace Laser Technology and System Department, Shanghai Institute of Optics and Fine Mechanics, Chinese Academy of Sciences, Shanghai 201800, China}\\
$^2${Shanghai Key Laboratory of All Solid-State Laser and Applied Techniques, Shanghai Institute of Optics and Fine Mechanics, Chinese Academy of Sciences, Shanghai 201800, China}\\
$^3${Center of Materials Science and Optoelectronics Engineering, University of Chinese Academy of Sciences, Beijing 100049, China}\\
$^4${Hangzhou Institute for Advanced Study, University of Chinese Academy of Sciences, Hangzhou 310024, China}\\
$^{*}$longwei20@mails.ucas.ac.cn\\
$^{*}$djchen@siom.ac.cn}

\begin{abstract}
\noindent We present a straightforward method to extend the noise-tracking bandwidth for self-correction algorithms in free-running dual-comb interferometry, leveraging coherent-harmonic-enhanced dual-comb spectroscopy. 
As a proof of concept, we employed both this novel architecture and a conventional one to perform free-running dual-comb spectroscopy of a $\text{H}^{13}\text{C}^{14}\text{N}$ gas cell, demonstrating a 20-fold increase in tracking bandwidth at the same spectral resolution of 12.5 MHz.
Since this approach improves the tracking bandwidth by generating harmonic centerbursts within an interferogram period, it decouples the tracking bandwidth from the repetition rate difference, thus avoiding spectral acquisition bandwidth narrowing. This significantly broadens the outlook for free-running dual-comb spectroscopy.

\end{abstract}

\maketitle

\medskip
\section{Introduction}

Dual-comb spectroscopy (DCS) is a powerful technique that allows instantaneous broadband, high-frequency accuracy, and high-resolution spectral acquisition \cite{DualcombSpectroscopy2016coddington}. However, this technique is extremely sensitive to the mutual coherence and stability of the dual combs owing to its physical principles: in the time domain, timing jitter or carrier-envelope phase (CEP) drift introduces phase noise into the interferogram (IGM) generated by asynchronous optical sampling, leading to spectral distortions and preventing coherent averaging over multiple periods to improve signal-to-noise ratio (SNR); in the frequency domain, tooth linewidth on the order of \~100 kHz for common free-running dual combs causes the tooth linewidth of the RF comb generated by multiheterodyne beating to significantly exceed its tooth spacing, resulting in a loss of spectral resolution.  Therefore, sophisticated locking systems, often incorporating the classic f-2f technique \cite{CarrierEnvelopePhase2000jones}, are employed to maintain mutual coherence and stability in typical implementations \cite{CoherentDualcomb2010coddington,PhasestableDualcomb2018chen}. Nevertheless, in practical implementations, especially in field deployments, these locking systems introduce high complexity and cost, as well as a lack of robustness.

Self-correction algorithms have been developed for free-running DCS as a post-treatment approach to retrieve mutual coherence between two OFCs over the past decade \cite{ComputationalMultiheterodyne2016burghoff,SelfcorrectedChipbased2017hébert,GeneralizedMethod2019burghoff,ComputationalDopplerlimited2019sterczewski} . This approach evaluates the relative phase or frequency noise by executing algorithms such as extended Kalman filter (EKF) \cite{ComputationalMultiheterodyne2016burghoff}, cross-relation and cross-ambiguity function (CRF and CAF) \cite{SelfcorrectedChipbased2017hébert,SelfCorrectionLimits2019hebert}, computational coherent averaging (CoCoA) \cite{ComputationalCoherent2019sterczewski,QuasirealtimeDualcomb2022tian}, short-time Fourier transform (STFT) \cite{DigitalError2019yu}, constant fraction discriminator (CFD) \cite{ComputationalDopplerlimited2019sterczewski} and so on \cite{FreerunningDualcomb2021yan,CoherentlyAveraged2023phillips} on the IGMs stream, and then compensates the noise through digital phase shift and resampling. Since all the jitter information is extracted from the IGMs, this method is free from sophisticated and costly locking systems, thereby offering a promising prospect of greatly simplifying DCS implementations.

    \begin{figure*}[t!]
        \centering
        \includegraphics[clip,width=1\linewidth]{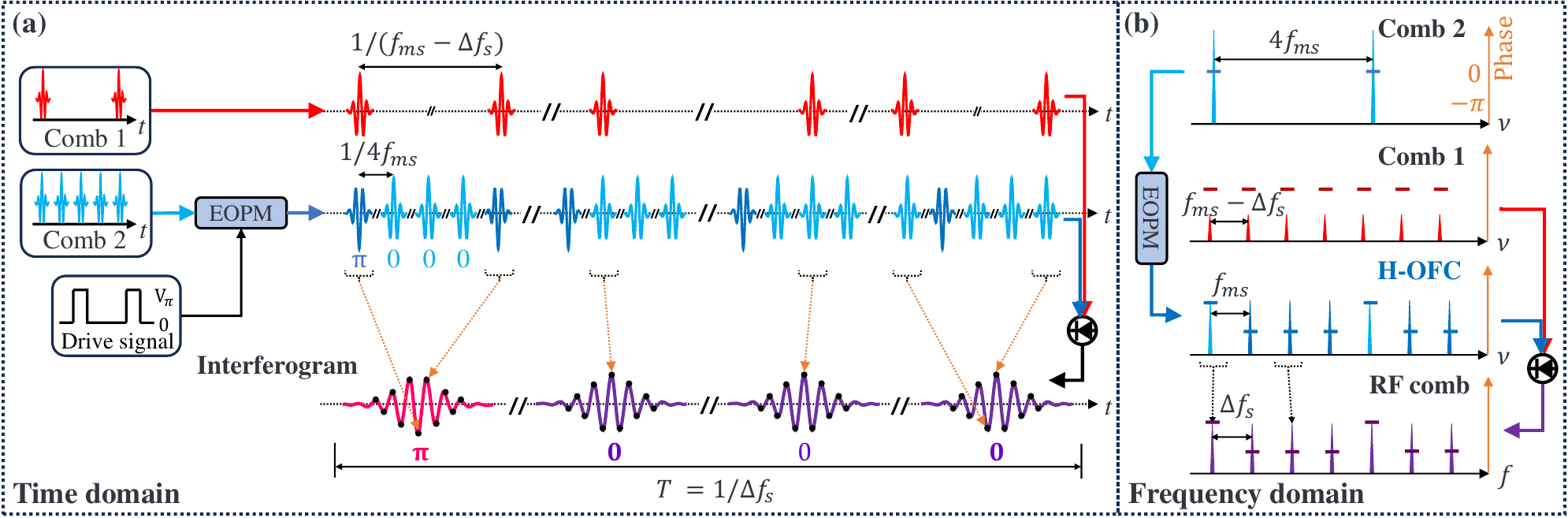}
        \caption{Concept of the architecture of CH-EDCS \cite{SpectralMode2025long} realized by electro-optic phase modulation. EOPM: electro-optic phase modulator, H-OFC: coherent-harmonic optical frequency comb.}
    \label{fig:concept}
    \end{figure*}

However, although the principles of each algorithm are different, they all face the same limitations in classic DCS using mode-locked optical frequency combs (OFCs) \cite{GeneralizedMethod2019burghoff,SelfCorrectionLimits2019hebert}. In these most common cases, the main energy per IGM is focused on a small region so-called "centerburst" while the other region is dominated by additive noise \cite{SensitivityCoherent2010newbury}. Therefore, the IGM provides only a single effective sampling within a period of $1/\Delta f_{s}$, where $\Delta f_{s}$ is the line spacing difference between the dual combs (which also corresponds to the repetition rate difference in conventional DCS). This inherently limits the phase-noise tracking speed to $\Delta f_{s}$ and a bandwidth of $\Delta f_{s}/2$ owing to Nyquist sampling theory \cite{SelfCorrectionLimits2019hebert}. In other words, only the partial phase noise band-limited to $\Delta f_{s}$ can be accurately estimated \cite{GeneralizedMethod2019burghoff, ComputationalDopplerlimited2019sterczewski}, and the remaining noise will be undersampled and disrupt the subsequent compensation. Moreover, when given the spectral resolution (e.g., $f_{rep}$), the $\Delta f_{s}$ and optical spectral bandwidth $\Delta \nu$ are in a trade-off relationship \cite{DualcombSpectroscopy2016coddington}, satisfying
\begin{equation}\label{eq:trade-off eq}
\Delta \nu \le \frac{f_{rep}^2}{2\Delta f_{s}} = \frac{f_{rep}^2}{4BW_{tracking}}, 
\end{equation}
where $\Delta f_{s}$ is the line spacing difference of the dual combs as well as the line spacing of the RF comb. This suggests that higher tracking bandwidths sometimes need to come at the expense of spectral bandwidth, as demonstrated in past work \cite{QuasirealtimeDualcomb2022tian}.


Recently, we reported a novel DCS concept, termed coherent-harmonic-enhanced dual-comb spectroscopy (CH-EDCS), to substitute conventional coherent averaging for signal-to-noise ratio (SNR) enhancement, realizing a reduction exceeding 300-fold in averaging time for comparable SNR in conventional DCS \cite{SpectralMode2025long}. 
Here, we demonstrate that this concept can effectively extend the noise-tracking bandwidth of the self-correction algorithms, without compromising spectral resolution or bandwidth. As a proof of concept, we employed two commercially available OFCs in the 1550-nm region and the cross-ambiguity-function-based self-correction algorithm, performing free-running dual comb spectroscopy of $\mathrm{H^{13}C^{14}N}$ with a 20-fold improvement in tracking bandwidth.

\medskip
\section{Principles and Setups}

\subsection{Extension of noise-tracking speed or bandwidth}

One architecture for this CH-EDCS, realized by electro-optic phase modulation, is depicted in Fig. \ref{fig:concept}. It employs two combs whose mode spacings (or repetition frequencies) have an approximate integer ratio. For instance, one operates at $4f_{ms}$ (Comb 2) and the other at $f_{ms}-\Delta f_s$ (Comb 1). Comb 2 is then phase modulated by an electro-optic phase modulator (EOPM) driven by a periodic signal of $[\pi, 0, 0, 0]$ (with a period exactly $1/f_{ms}$), and is transformed into a coherent-harmonic OFC (H-OFC). The subsequent asynchronous optical sampling of the dual combs produces an IGM with the same coherent-harmonic characteristic: multiple centerbursts possessing a specific CEP profile of $[\pi, 0, 0, 0]$ within a period of $1/\Delta f_s$. In the frequency domain, Comb 2 is densified \cite{SpectralSelfimaging2011caraquitena,ReconfigurableMultiwavelength2011beltran} to form a new comb with a mode spacing of $f_{ms}$, which beats with the comb 1, yielding a radio-frequency (RF) comb with a mode spacing of $\Delta f_s$ (more detailed principles of this architecture can be found in our previous work \cite{SpectralMode2025long}).

The ambiguity-function-based self-correction algorithm measures the similarity between consecutive centerbursts and the initial one, extracting the noise information including timing jitter, phase fluctuations and carrier-frequency drifts \cite{SelfcorrectedChipbased2017hébert}. Although the centerbursts are discrete, consecutive correction covered the whole optical path difference is needed for treatment for free-induction decay (FID) beyond the centerbursts. Therefore, this information is interpolated to match the data sampling rate of the IGMs. 
In this architecture, since the ideal centerbursts (with no phase noise) exhibit a periodic phase profile, the observed phase fluctuation need to be compared to this profile first to obtain the actual drifts (e.g., deviations from the ideal $[\pi, 0, 0, 0]$ profile). For the more general case, this phase profile is determined by the signals driving the modulators \cite{SpectralMode2025long}. Most importantly, the noise tracking speed and bandwidth are increased to $m\cdot\Delta f_{s}$ and $m\cdot\Delta f_{s}/2$, respectively, where $m \in \mathbb{Z}^+$. Therefore, the trade-off relationship is updated to: 
\begin{equation}\label{eq:trade-off eq2}
\Delta \nu \le \frac{f_{s}^2}{2\Delta f_{s}} = m\cdot\frac{f_{s}^2}{4BW_{tracking}}, 
\end{equation}
or, equivalently: 
\begin{equation}\label{eq:BW tracking limit}
BW_{tracking} \le  m\cdot\frac{f_{s}^2}{4\Delta \nu}, 
\end{equation}
indicating an \textit{m}-fold relaxation of the constraint.

\subsection{Experimental Setups} 

    \begin{figure*}[!ht]
        \centering
        \includegraphics[clip,width=0.85\linewidth]{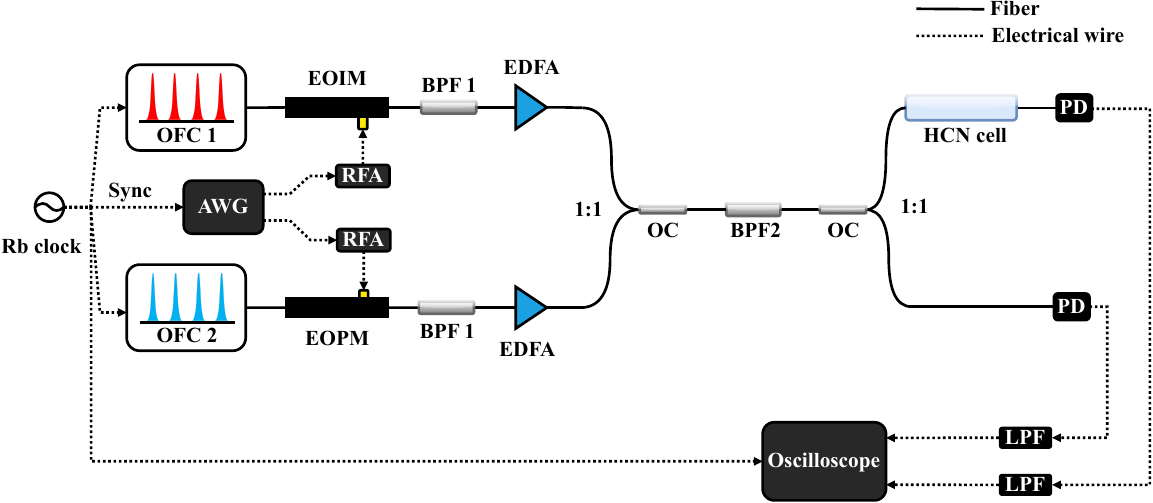}
        \caption{Proof-of-concept experimental setup. OFC: optical frequency comb, EOPM: electro-optic phase modulator, EOIM:electro-optic intensity modulator, AWG: arbitrary waveform generator, RFA: RF amplifier, BPF: band-pass filter, EDFA: erbium-doped fiber amplifier, OC: optical coupler, PD: photodetector, LPF: low-pass filter.}
    \label{fig:E-Setup}
    \end{figure*}

To validate this method, we conducted a proof-of-concept experiment with a $\mathrm{H^{13}C^{14}N}$ gas cell using both conventional architecture and this new architecture. All measurements were performed at the same spectral bandwidth $\Delta \nu$ of $\sim$ 0.3 nm (at 12 dB, set by the optical band-pass filter) and a spectral resolution $f_{rep}$ of 12.5 MHz. To avoid IGM spectral aliasing (Eq. \ref{eq:trade-off eq}) and ensure there are integer points within each IGM period, the $\Delta f_{s}$ was set to 1 kHz, leading to an IGM acquisition speed ($1/\Delta f_{s}$) of 1 ms.

The experimental setup is shown in Fig. \ref{fig:E-Setup}. The two OFCs were loosely locked to an RF reference for timing synchronization between pulses and phase modulation, which provided little benefit to the coherence required by the DCS \cite{DualcombSpectroscopy2016coddington} because of the leverage factor \textit{N} from RF to optical frequency embedded in the "comb equation" $\nu = N\cdot f_{rep}+f_{CEO}$. OFC 1 was pulse-picked by an electro-optic intensity modulator (EOIM), enabling it to emulate a low-repetition-rate OFC at 12.5 MHz. Replacing the electro-optic phase modulator (EOPM) connected after OFC 2 with another EIOM allows for the implementation of the conventional architecture. The remainder of the experimental setup in this study is detailed in our previous work \cite{SpectralMode2025long}.

\section{Results}

\subsection{Noise Estimating}
\begin{figure}[!htb]
    \captionsetup[subfigure]{labelformat=empty}
        \centering
        \begin{subfigure}{0.9\linewidth}
          \includegraphics[width=\textwidth]{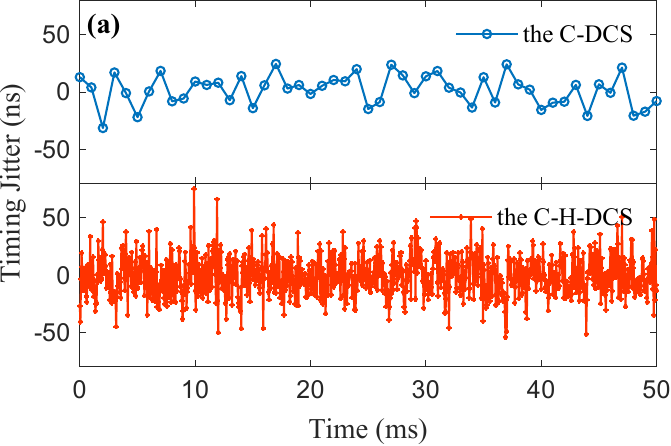}
        \end{subfigure}%
        \\
        \centering
        \begin{subfigure}{0.9\linewidth}
          \includegraphics[width=\linewidth]{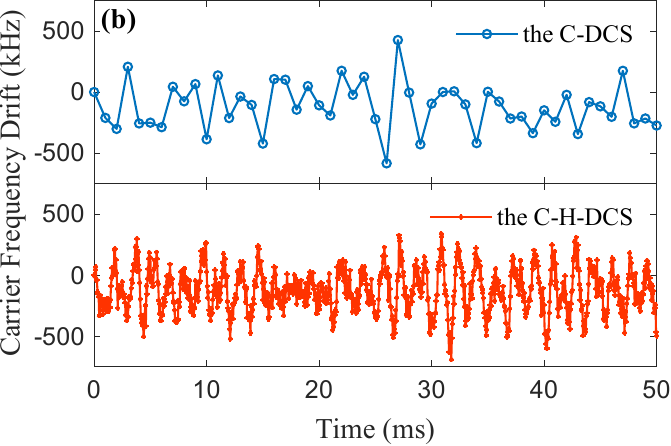}
        \end{subfigure}%
    \caption{(a) Timing jitter and (b) carrier frequency drift of the interferogram centerburst of the two architectures.}
    \label{fig:tj & cf}
    \end{figure}

Fig. \ref{fig:tj & cf} illustrates the estimated timing jitter and carrier frequency drift of the IGM centerbursts for both architectures within a 50 ms window, as determined by the cross-ambiguity function. For both architectures, the carrier frequency drift was on the order of 1 MHz, consistent with the typical linewidth of these optical frequency combs in a free-running state. Moreover, the C-H-DCS trace exhibited more detailed jitter and provided sufficient sampling to capture these fluctuation details. In contrast, the C-DCS showed insufficient sampling of carrier frequency fluctuations, with only the initial and final sampling points visible for some fluctuation details.

\begin{figure}[!htb]
    \captionsetup[subfigure]{labelformat=empty}
        \centering
        \begin{subfigure}{1\linewidth}
          \includegraphics[width=\textwidth]{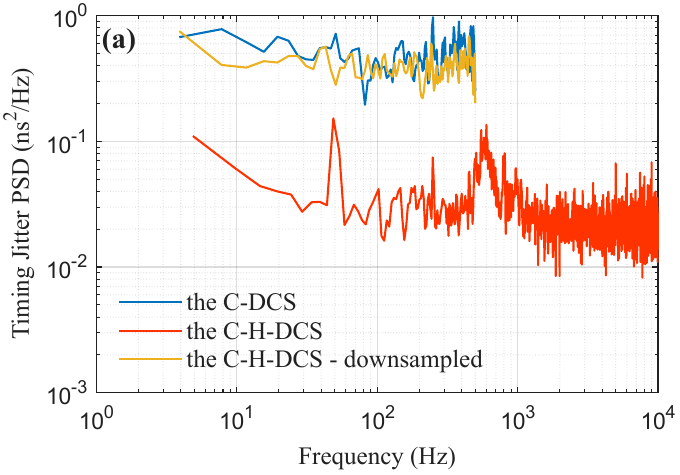}
        \end{subfigure}%
        \\
        \centering
        \begin{subfigure}{1\linewidth} 
          \includegraphics[width=\linewidth]{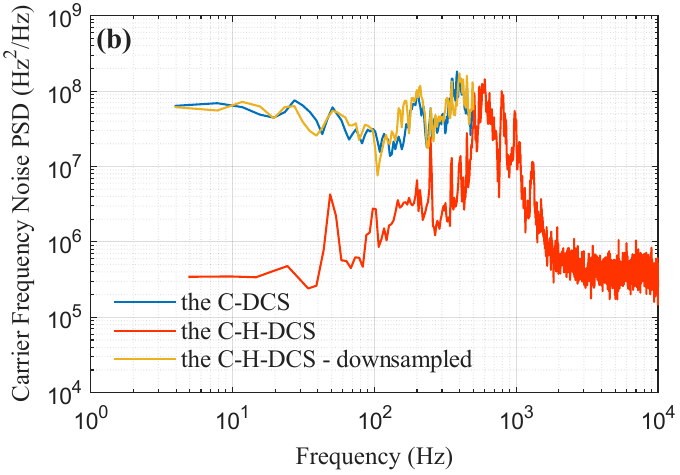}
        \end{subfigure}%
    \caption{Comparison of the PSD of the estimated (a) timing jitter and (b) carrier frequency drift. The yellow traces correspond to PSDs calculated from the noise information of the C-H-DCS after downsampling by a factor of 20 in the time domain.}
    \label{fig:PSD}
    \end{figure}

The power spectral densities (PSDs) of the timing jitter and the carrier frequency drift are calculated to verify the tracking bandwidth, as shown in Fig. \ref{fig:PSD}. The PSDs of the C-H-DCS clearly reveal noise information below 10 kHz, particularly showing noise peaks between 0.5 and 1 kHz. In contrast, the C-DCS PSDs show a bandwidth limited to 0.5 kHz, failing to display the noise peaks and exhibiting a higher overall level. We attribute this higher level to the "folding" of high-frequency noise (above 0.5 kHz) due to undersampling. This is further supported by the PSDs of the downsampled C-H-DCS, which show a high degree of agreement with the PSDs of the C-DCS.

\subsection{Corrected Interferograms and Spectra}
\begin{figure*}[!htb]
    \captionsetup[subfigure]{labelformat=empty}
        \centering
        \begin{subfigure}{0.9\textwidth}
          \includegraphics[width=\textwidth]{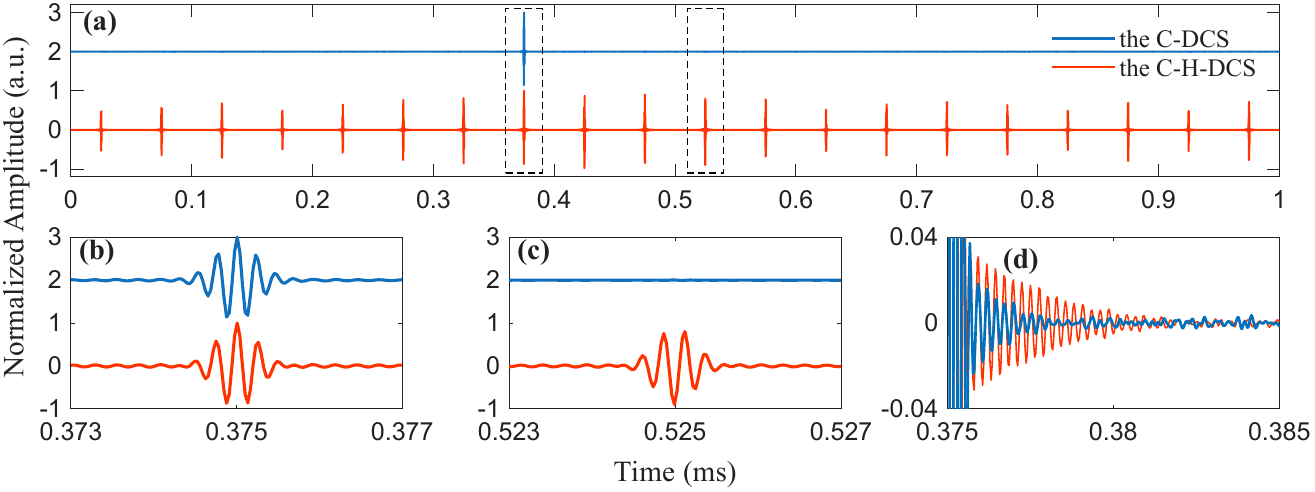}
        \end{subfigure}%
        \\
        \centering
        \begin{subfigure}{0.9\textwidth}
          \includegraphics[width=\textwidth]{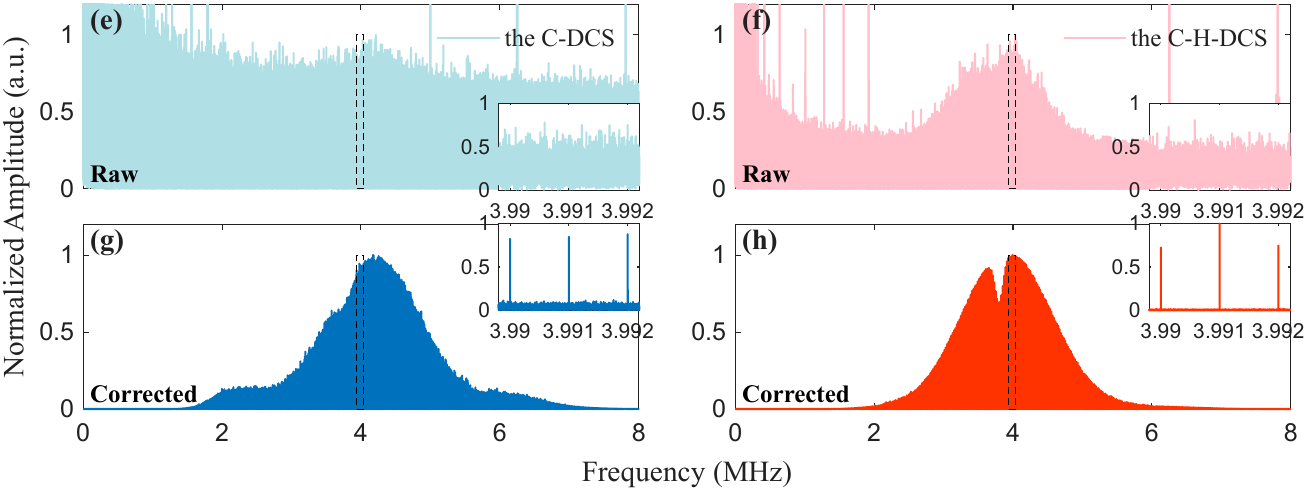}
        \end{subfigure}%
    \caption{Averaged interferograms (IGMs) and spectra of the C-DCS and the C-H-DCS after correction. 
    (a) Averaged IGMs within a 1 ms period. The centerburst repetition rate of the C-H-DCS is 20 times that of the C-DCS. 
    (b-c) Zoom of the centerbursts of two IGMs. The centerburst of the C-H-DCS in (b) shows a inverse phase compared to that in (c). 
    (d) Overlap of the Free-Induction Decay (FID) portions of the two IGMs. 
    (e-f) Spectra of the raw IGMs before correction. The C-H-DCS shows a higher signal-to-noise ratio (SNR) due to the spectral-mode enhancement mechanism of the multiple centerbursts \cite{SpectralMode2025long}. 
    (g-h) Spectra of the IGMs after correction. The two spectra show the same line spacing of 1 kHz. The unflatness of the comb lines as illustrated in the zoom is derived from limited extinction ratio of the electro-optic intensity modulator (EOIM) used for pulse picking \cite{TunableResolution2016vieira}.}
    \label{fig:Interferograms & Spectra}
    \end{figure*}

Figures \ref{fig:Interferograms & Spectra} (a-d) and (g-h) show the IGMs and spectra, respectively, obtained by coherent averaging and Fourier transforming the corrected IGMs of the two architectures over 1.999 s. After coherent averaging, the C-H-DCS exhibits the same centerburst shape as the C-DCS, but with a longer FID duration, as shown in Fig. \ref{fig:Interferograms & Spectra} (d). Given that this is the same absorption peak and the FID duration is determined solely by the absorption peak linewidth, the longer duration observed in the averaged C-H-DCS IGM suggests improved preservation of its FID waveform's coherence for larger optical path differences (OPDs) during the coherent averaging process, and implies a longer OPD correction range.

In the frequency domain, In contrast to C-DCS setups that increase the repetition rate difference from 1 kHz to 20 kHz, thereby broadening the RF comb's line spacing and spectral width, the C-H-DCS preserves the same RF spectral bandwidth as the C-DCS. This characteristic helps avoid encountering 0 Hz or $f_s/2$ frequencies, thereby preventing aliasing. Both corrected spectra clearly display a 1 kHz comb line spacing, which was not visible in the raw spectra. The C-H-DCS spectrum exhibited narrower and deeper absorption features compared to the C-DCS, which corresponds to the longer FID duration observed  in the time domain. Moreover, the higher SNR of the C-H-DCS spectrum can be attributed to the additional centerbursts, which theoretically lead to a $\sqrt{10}$-fold mode amplitude enhancement \cite{SpectralMode2025long}.

\begin{figure}[!ht]
        \centering
        \includegraphics[clip,width=1\linewidth]{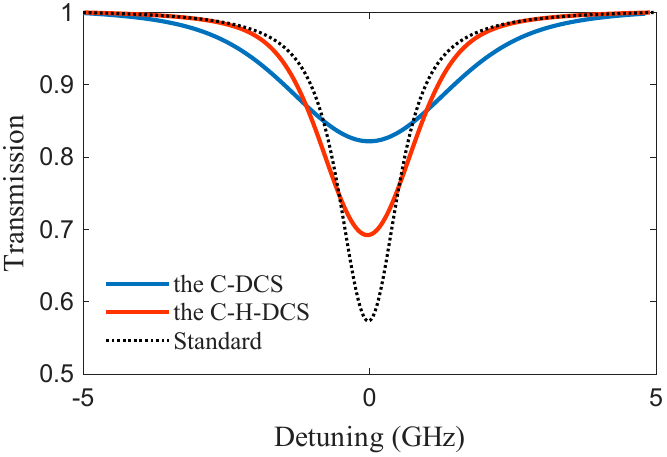}
        \caption{Weighted Voigt profile fits of the P7 absorption line of $\text{H}^{13}\text{C}^{14}\text{N}$ measured by the C-DCS and the C-H-DCS. The black dashed line represents a standard measurement obtained using a continuous-wave laser as an intermediate oscillator \cite{OpticalReferencing2010deschênes,FreerunningOperation2020guay,ModeresolvedDualcomb2021yu}.}
    \label{fig:fits}
    \end{figure}

The fits of the absorption line for the two architectures are shown in Fig. \ref{fig:fits}. The C-H-DCS architecture effectively improves the correction outcome, manifesting as a deeper and narrower profile that more closely matches the standard. This result aligns with our previous work on free-running dual-comb spectroscopy with varying interferogram refresh rates \cite{SelfCorrectionFreeRunning2025maa}, where we used two similar 9-figure cavity fiber mode-locked OFCs and increased the $\Delta f_r$ from approximately 1 kHz to about 80 kHz by directly adjusting the cavity length difference. The remaining discrepancy between the C-H-DCS fit and the standard is attributed to still insufficient tracking bandwidth. While further increasing the bandwidth could improve this result, it was limited in this proof-of-concept experiment by the repetition rate of the commercial optical combs we employed.

\medskip
\section{Discussion}

\subsection{Generalizability to other algorithms}

As the noise-tracking-bandwidth extension method introduced in this work is a hardware-level improvement, it directly multiplies the scanning rate of the IGM's centerburst without changing spectral resolution. Therefore, we believe this method is broadly suitable for other existing self-correction algorithms. 

For example, in methods that estimate the phase noise by sampling discrete centerbursts, such as the Short-Time Fourier Transform (STFT) and the Constant Fraction Discriminator (CFD), the multiplication of the centerburst's scanning rate will certainly improve the tracking bandwidth. Meanwhile, in methods that continuously track the interferogram, such as the Extended Kalman Filter (EKF) and Computational Coherent Averaging (CoCoA), the increased number of centerbursts can significantly improve estimator accuracy and SNR relative to noise, producing a similar effect.

Limited by the repetition rate of our OFCs (where a higher tracking bandwidth could lead to insufficient spectral bandwidth), we did not extend the tracking bandwidth to the level required for perfect correction in this proof-of-concept study. However, using higher repetition-rate light sources in future work would allow for larger \textit{m} and $f_s$, enabling the tracking bandwidth to reach the approximately predicted level by previous work \cite{SelfCorrectionLimits2019hebert}.

\medskip
\subsection{Free-induction decay in self-correction method}

It is important to emphasize that self-correction methods should primarily focus on the free-induction decay (FID) rather than solely on the centerbursts. While the centerburst has a short duration and serves as the primary region for noise information extraction, making it relatively straightforward to treat, the crucial information regarding narrow absorption peaks resides not within the centerburst but within the longer-duration FID, for correcting which the noise information is basically derived from interpolating the original sampling points. Consequently, effectively treating the FID in a self-correction method requires a higher tracking bandwidth. Therefore, a demonstration based solely on RF-comb linewidth without considering the absorption peak lacks practical significance.

A potential concern in this process is that the noise information from the centerbursts is used to correct the FID, yet the noise characteristics of these two sections may not be entirely identical. This concern may be more pronounced in this new architecture when the H-OFC's $f_{rep}$ surpasses the absorption peak linewidth (a situation that can arise as the DCS resolution is then $f_{rep}/m$ rather than $f_{rep}$). The FID from a preceding centerburst in the interferogram may overlap with the subsequent centerburst, leading to correction of this portion of the FID using phase noise information from the latter. However, since the FID excited by the preceding pulse and the subsequent pulse propagate along a common path, they are subjected to the same common-path noise. Consequently, any residual noise difference originates almost purely from the intrinsic noise within the resonant cavity between the two adjacent pulses. This corresponds to phase noise in the hundreds of MHz to GHz range, and is typically very low and therefore negligible in practical comb sources \cite{kimUltralownoiseModelockedFiber2016}.

\medskip
\section{Conclusion}

In conclusion, we showed that the noise-tracking speed or bandwidth in free-running dual-comb interferometry can be efficiently multiplied by employing a new architecture that substitutes a comb with a higher-repetition-rate H-OFC. 
The multiplication factor, m, is the harmonic number of the H-OFC, defined as the ratio of its repetition rate to its mode spacing. 
We validated our method by measuring an absorption peak around 1550 nm of an $\mathrm{H^{13}C^{14}N}$ gas cell using both a conventional architecture and this new architecture. The results demonstrated an effective improvement in the correction outcome, aligning with results from our previous work regarding varying tracking bandwidths in the conventional architecture. 
This method is of great significance for free-running DCS as it provides a straightforward relaxation of the trade-off among spectral resolution, tracking bandwidth, and optical bandwidth. Thus, the perfect self-correction with ideal tracking bandwidth predicted by previous work \cite{SelfCorrectionLimits2019hebert} could become much easier to access.

\medskip
\noindent\textbf{Acknowledgments}

\begin{footnotesize}
\noindent This work was supported by National Key Research and Development Program of China under Grant2020YFC2200300. The authors thank Prof. Tang Li for providing the Menlo-250-WG-PM, Yatan Xiong for lending the EDFA, and Bowen Ruan for his assistance with certain details.
\end{footnotesize}
\medskip

\noindent\textbf{Disclosures}

\begin{footnotesize}
\noindent The authors declare no conflicts of interest.
\end{footnotesize}
\medskip

\noindent\textbf{Data availability} 

\begin{footnotesize}
\noindent Data underlying the results presented in this paper are not publicly available at this time but may be obtained from the authors upon reasonable request.
\end{footnotesize}
\medskip

\medskip
\bibliography{ref}

\begin{thebibliography}{10}
\expandafter\ifx\csname url\endcsname\relax
  \def\url#1{\texttt{#1}}\fi
\expandafter\ifx\csname urlprefix\endcsname\relax\def\urlprefix{URL }\fi
\providecommand{\bibinfo}[2]{#2}
\providecommand{\eprint}[2][]{\url{#2}}

\bibitem{DualcombSpectroscopy2016coddington}
\bibinfo{author}{Coddington, I.}, \bibinfo{author}{Newbury, N.} \& \bibinfo{author}{Swann, W.}
\newblock \bibinfo{title}{Dual-comb spectroscopy}.
\newblock \emph{\bibinfo{journal}{Optica}} \textbf{\bibinfo{volume}{3}}, \bibinfo{pages}{414--426} (\bibinfo{year}{2016}).

\bibitem{CarrierEnvelopePhase2000jones}
\bibinfo{author}{Jones, D.~J.}
\newblock \bibinfo{title}{Carrier-{{Envelope Phase Control}} of {{Femtosecond Mode-Locked Lasers}} and {{Direct Optical Frequency Synthesis}}}.
\newblock \emph{\bibinfo{journal}{Science}} \textbf{\bibinfo{volume}{288}}, \bibinfo{pages}{635--639} (\bibinfo{year}{2000}).

\bibitem{CoherentDualcomb2010coddington}
\bibinfo{author}{Coddington, I.}, \bibinfo{author}{Swann, W.} \& \bibinfo{author}{Newbury, N.}
\newblock \bibinfo{title}{Coherent dual-comb spectroscopy at high signal-to-noise ratio}.
\newblock \emph{\bibinfo{journal}{Physical Review A}} \textbf{\bibinfo{volume}{82}}, \bibinfo{pages}{043817} (\bibinfo{year}{2010}).

\bibitem{PhasestableDualcomb2018chen}
\bibinfo{author}{Chen, Z.}, \bibinfo{author}{Yan, M.}, \bibinfo{author}{H{\"a}nsch, T.~W.} \& \bibinfo{author}{Picqu{\'e}, N.}
\newblock \bibinfo{title}{A phase-stable dual-comb interferometer}.
\newblock \emph{\bibinfo{journal}{Nature Communications}} \textbf{\bibinfo{volume}{9}}, \bibinfo{pages}{3035} (\bibinfo{year}{2018}).

\bibitem{ComputationalMultiheterodyne2016burghoff}
\bibinfo{author}{Burghoff, D.}, \bibinfo{author}{Yang, Y.} \& \bibinfo{author}{Hu, Q.}
\newblock \bibinfo{title}{Computational multiheterodyne spectroscopy}.
\newblock \emph{\bibinfo{journal}{Science Advances}} \textbf{\bibinfo{volume}{2}}, \bibinfo{pages}{e1601227} (\bibinfo{year}{2016}).

\bibitem{SelfcorrectedChipbased2017hébert}
\bibinfo{author}{H{\'e}bert, N.~B.} \emph{et~al.}
\newblock \bibinfo{title}{Self-corrected chip-based dual-comb spectrometer}.
\newblock \emph{\bibinfo{journal}{Optics Express}} \textbf{\bibinfo{volume}{25}}, \bibinfo{pages}{8168} (\bibinfo{year}{2017}).

\bibitem{GeneralizedMethod2019burghoff}
\bibinfo{author}{Burghoff, D.}, \bibinfo{author}{Han, N.} \& \bibinfo{author}{Shin, J.~H.}
\newblock \bibinfo{title}{Generalized method for the computational phase correction of arbitrary dual comb signals}.
\newblock \emph{\bibinfo{journal}{Optics Letters}} \textbf{\bibinfo{volume}{44}}, \bibinfo{pages}{2966} (\bibinfo{year}{2019}).

\bibitem{ComputationalDopplerlimited2019sterczewski}
\bibinfo{author}{Sterczewski, {\L}.~A.}, \bibinfo{author}{Przew{\l}oka, A.}, \bibinfo{author}{Kaszub, W.} \& \bibinfo{author}{Sotor, J.}
\newblock \bibinfo{title}{Computational {{Doppler-limited}} dual-comb spectroscopy with a free-running all-fiber laser}.
\newblock \emph{\bibinfo{journal}{APL Photonics}} \textbf{\bibinfo{volume}{4}}, \bibinfo{pages}{116102} (\bibinfo{year}{2019}).

\bibitem{SelfCorrectionLimits2019hebert}
\bibinfo{author}{Hebert, N.~B.}, \bibinfo{author}{{Michaud-Belleau}, V.}, \bibinfo{author}{Deschenes, J.-D.} \& \bibinfo{author}{Genest, J.}
\newblock \bibinfo{title}{Self-{{Correction Limits}} in {{Dual-Comb Interferometry}}}.
\newblock \emph{\bibinfo{journal}{IEEE Journal of Quantum Electronics}} \textbf{\bibinfo{volume}{55}}, \bibinfo{pages}{1--11} (\bibinfo{year}{2019}).

\bibitem{ComputationalCoherent2019sterczewski}
\bibinfo{author}{Sterczewski, L.~A.}, \bibinfo{author}{Westberg, J.} \& \bibinfo{author}{Wysocki, G.}
\newblock \bibinfo{title}{Computational coherent averaging for free-running dual-comb spectroscopy}.
\newblock \emph{\bibinfo{journal}{Optics Express}} \textbf{\bibinfo{volume}{27}}, \bibinfo{pages}{23875} (\bibinfo{year}{2019}).

\bibitem{QuasirealtimeDualcomb2022tian}
\bibinfo{author}{Tian, H.} \emph{et~al.}
\newblock \bibinfo{title}{Quasi-real-time dual-comb spectroscopy with 750-{{MHz Yb}}:fiber combs}.
\newblock \emph{\bibinfo{journal}{Optics Express}} \textbf{\bibinfo{volume}{30}}, \bibinfo{pages}{28427} (\bibinfo{year}{2022}).

\bibitem{DigitalError2019yu}
\bibinfo{author}{Yu, H.} \emph{et~al.}
\newblock \bibinfo{title}{Digital error correction of dual-comb interferometer without external optical referencing information}.
\newblock \emph{\bibinfo{journal}{Optics Express}} \textbf{\bibinfo{volume}{27}}, \bibinfo{pages}{29425} (\bibinfo{year}{2019}).

\bibitem{FreerunningDualcomb2021yan}
\bibinfo{author}{Yan, Q.} \emph{et~al.}
\newblock \bibinfo{title}{A free-running dual-comb spectrometer with intelligent temporal alignment algorithm}.
\newblock \emph{\bibinfo{journal}{Optics \& Laser Technology}} \textbf{\bibinfo{volume}{141}}, \bibinfo{pages}{107175} (\bibinfo{year}{2021}).

\bibitem{CoherentlyAveraged2023phillips}
\bibinfo{author}{Phillips, C.~R.} \emph{et~al.}
\newblock \bibinfo{title}{Coherently averaged dual-comb spectroscopy with a low-noise and high-power free-running gigahertz dual-comb laser}.
\newblock \emph{\bibinfo{journal}{Optics Express}} \textbf{\bibinfo{volume}{31}}, \bibinfo{pages}{7103} (\bibinfo{year}{2023}).

\bibitem{SensitivityCoherent2010newbury}
\bibinfo{author}{Newbury, N.~R.}, \bibinfo{author}{Coddington, I.} \& \bibinfo{author}{Swann, W.}
\newblock \bibinfo{title}{Sensitivity of coherent dual-comb spectroscopy}.
\newblock \emph{\bibinfo{journal}{Optics Express}} \textbf{\bibinfo{volume}{18}}, \bibinfo{pages}{7929--7945} (\bibinfo{year}{2010}).

\bibitem{SpectralMode2025long}
\bibinfo{author}{Long, W.} \emph{et~al.}
\newblock \bibinfo{title}{Spectral {{Mode Enhancement}} in {{Coherent-harmonic Dual-comb Spectroscopy Enables Exceeding}} 300-fold {{Averaging Time Reduction}}} (\bibinfo{year}{2025}).
\newblock \eprint{arXiv:2504.10426}.

\bibitem{SpectralSelfimaging2011caraquitena}
\bibinfo{author}{Caraquitena, J.}, \bibinfo{author}{Beltr{\'a}n, M.}, \bibinfo{author}{Llorente, R.}, \bibinfo{author}{Mart{\'i}, J.} \& \bibinfo{author}{Muriel, M.~A.}
\newblock \bibinfo{title}{Spectral self-imaging effect by time-domain multilevel phase modulation of a periodic pulse train}.
\newblock \emph{\bibinfo{journal}{Optics Letters}} \textbf{\bibinfo{volume}{36}}, \bibinfo{pages}{858} (\bibinfo{year}{2011}).

\bibitem{ReconfigurableMultiwavelength2011beltran}
\bibinfo{author}{Beltran, M.}, \bibinfo{author}{Caraquitena, J.}, \bibinfo{author}{Llorente, R.} \& \bibinfo{author}{Marti, J.}
\newblock \bibinfo{title}{Reconfigurable {{Multiwavelength Source Based}} on {{Electrooptic Phase Modulation}} of a {{Pulsed Laser}}}.
\newblock \emph{\bibinfo{journal}{IEEE Photonics Technology Letters}} \textbf{\bibinfo{volume}{23}}, \bibinfo{pages}{1175--1177} (\bibinfo{year}{2011}).

\bibitem{TunableResolution2016vieira}
\bibinfo{author}{Vieira, F.~S.}, \bibinfo{author}{Cruz, F.~C.}, \bibinfo{author}{Plusquellic, D.~F.} \& \bibinfo{author}{Diddams, S.~A.}
\newblock \bibinfo{title}{Tunable resolution terahertz dual frequency comb spectrometer}.
\newblock \emph{\bibinfo{journal}{Optics Express}} \textbf{\bibinfo{volume}{24}}, \bibinfo{pages}{30100} (\bibinfo{year}{2016}).

\bibitem{SelfCorrectionFreeRunning2025maa}
\bibinfo{author}{Ma, X.} \emph{et~al.}
\newblock \bibinfo{title}{Self-{{Correction}} in {{Free-Running Dual-Comb Spectroscopy With Varying Interferogram Refresh Rates}}}.
\newblock \emph{\bibinfo{journal}{IEEE Photonics Journal}} \textbf{\bibinfo{volume}{17}}, \bibinfo{pages}{1--5} (\bibinfo{year}{2025}).

\bibitem{OpticalReferencing2010deschênes}
\bibinfo{author}{Desch{\^e}nes, J.-D.}, \bibinfo{author}{Giaccarri, P.} \& \bibinfo{author}{Genest, J.}
\newblock \bibinfo{title}{Optical referencing technique with {{CW}} lasers as intermediate oscillators for continuous full delay range frequency comb interferometry}.
\newblock \emph{\bibinfo{journal}{Optics Express}} \textbf{\bibinfo{volume}{18}}, \bibinfo{pages}{23358} (\bibinfo{year}{2010}).

\bibitem{FreerunningOperation2020guay}
\bibinfo{author}{Guay, P.} \emph{et~al.}
\newblock \bibinfo{title}{Toward free-running operation of dual-comb fiber lasers for methane sensing}.
\newblock \emph{\bibinfo{journal}{Applied Optics}} \textbf{\bibinfo{volume}{59}}, \bibinfo{pages}{B35} (\bibinfo{year}{2020}).

\bibitem{ModeresolvedDualcomb2021yu}
\bibinfo{author}{Yu, H.}, \bibinfo{author}{Zhou, Q.}, \bibinfo{author}{Li, X.}, \bibinfo{author}{Wang, X.} \& \bibinfo{author}{Ni, K.}
\newblock \bibinfo{title}{Mode-resolved dual-comb spectroscopy using error correction based on single optical intermedium}.
\newblock \emph{\bibinfo{journal}{Optics Express}} \textbf{\bibinfo{volume}{29}}, \bibinfo{pages}{6271--6281} (\bibinfo{year}{2021}).

\bibitem{kimUltralownoiseModelockedFiber2016}
\bibinfo{author}{Kim, J.} \& \bibinfo{author}{Song, Y.}
\newblock \bibinfo{title}{Ultralow-noise mode-locked fiber lasers and frequency combs: Principles, status, and applications}.
\newblock \emph{\bibinfo{journal}{Advances in Optics and Photonics}} \textbf{\bibinfo{volume}{8}}, \bibinfo{pages}{465} (\bibinfo{year}{2016}).

\end{thebibliography}

\end{document}